\documentclass[sigconf]{acmart}

\usepackage{booktabs} 

\usepackage{booktabs} 
\usepackage{algorithm}
\usepackage{algorithmic}
\usepackage{color}
\usepackage{amsmath}
\usepackage{amssymb}
\usepackage{subfigure}
\usepackage{graphicx}
\usepackage{epstopdf}
\usepackage{multirow}
\usepackage{booktabs} 

\usepackage{lineno}
\usepackage{multirow}

\copyrightyear{2018}
\acmYear{2018}
\setcopyright{acmcopyright}
\acmConference[KDD 2018]{24th ACM SIGKDD International Conference on Knowledge Discovery \& Data Mining}{August 19--23, 2018}{London, United Kingdom}
\acmBooktitle{KDD 2018: 24th ACM SIGKDD International Conference on Knowledge Discovery \& Data Mining, August 19--23, 2018, London, United Kingdom}
\acmPrice{15.00}
\acmDOI{10.1145/3219819.3219865}
\acmISBN{978-1-4503-5552-0/18/08}

\begin{document}
\title{Inferring Metapopulation Propagation Network for Intra-city Epidemic Control and Prevention}

\author{Jingyuan Wang, Xiaojian Wang, Junjie Wu}
\affiliation{%
  \institution{Beihang University, Beijing, China}
}\email{{jywang, wangxj, wujj}@buaa.edu.cn}


\begin{abstract}
	Since the 21st century, the global outbreaks of infectious diseases such as SARS in 2003, H1N1 in 2009, and H7N9 in 2013, have become the critical threat to the public health and a hunting nightmare to the government. Understanding the propagation in large-scale metapopulations and predicting the future outbreaks thus become crucially important for epidemic control and prevention. In the literature, there have been a bulk of studies on modeling intra-city epidemic propagation but with the single population assumption (homogeneity). Some recent works on metapopulation propagation, however, focus on finding specific human mobility physical networks to approximate diseases transmission networks, whose generality to fit different diseases cannot be guaranteed. In this paper, we argue that the intra-city epidemic propagation should be modeled on a metapopulation base, and propose a two-step method for this purpose. The first step is to understand the propagation system by inferring the underlying disease infection network. To this end, we propose a novel network inference model called D$^2$PRI, which reduces the individual network into a sub-population network without information loss, and incorporates the power-law distribution prior and data prior for better performance. The second step is to predict the disease propagation by extending the classic SIR model to a metapopulation SIR model that allows visitors transmission between any two sub-populations. The validity of our model is testified on a real-life clinical report data set about the airborne disease in the Shenzhen city, China. The D$^2$PRI model with the extended SIR model exhibit superior performance in various tasks including network inference, infection prediction and outbreaks simulation. 
\end{abstract}

%
%



\maketitle


\section{Introduction}
Infectious diseases are serious threats to human life and health. From the Black Death resulting in about 75 million deaths in 1340s, to the 2017 outbreak of H3N2 influenza in Hongkong killing over 300 residents in just two months, the war between human beings and infectious diseases will never end. On the other hand, the developed transportation systems nowadays make long distance travel very convenient. Likewise, with mobility of infected persons, pathogens can be spread to large geographic space within a short period of time. The recent global epidemic outbreaks, including SARS in 2003~\cite{sars}, H1N1 in 2009~\cite{h1n1} and H7N9 in 2013~\cite{h7n9}, all have close relationship with transnational human mobilities. Understanding large spatial diseases transmission with human mobility and predicting outbreak process of epidemics in early stages, have become crucial problems in epidemic control and prevention.


\begin{figure}
	\centering
	\includegraphics[width=0.75\columnwidth]{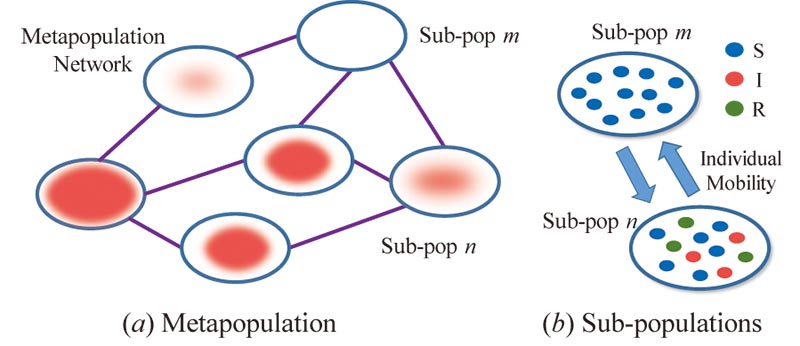}\\
	\caption{Illustration of the metapopulation SIR model.~\cite{metapopulation}}
	\label{fig:metapopulation}
\end{figure}

In the literature, many epidemic models have been proposed to reveal propagation dynamics of disease in different structures of population, such as the compartment models~\cite{sir} for the small size and individual ``well-mixed'' population, and network epidemiology models~\cite{metapopulation} for individuals with complex contact relationship in a single population. For epidemic propagation in a large-scale spatial area, the most widely used model is the {\em metapopulation} model. A meta-population refers to a group of separated sub-populations of the same species which are connected by an interaction network. Large-scale epidemic outbreaks, such as global transmission of influenzas, can be modeled as a propagation of pathogens through a metapopulation network, in which cities of different countries are modeled as sub-populations and inter-city human mobility are modeled as the network connecting the sub-populations (see Fig.~\ref{fig:metapopulation}).

The metapopulation model has achieved great success in empirical large scale epidemic propagation studies. For example, the studies~\cite{sars,brockmann2013hidden} use the worldwide aviation network to analyze the propagation of SARS and H1N1 in the global city metapopulation, while the study~\cite{malaria} uses a cell-phone user mobility network to analyze Malaria propagation in an inter-settlement metapopulation of Kenya. However, because obtaining detailed mobility data of all cities in the world or even in a country is often practically impossible, most of these works can only build epidemic propagation networks at the inter-city level using coarse-grained mobility data, assuming that all contacts and infections between individuals in the same city are homogeneous. With the rapid development of metropolis in the worldwide, social structures inside a city also become more and more complex, and therefore the homogeneous mixture assumption of intra-city population no longer holds. Moreover, it is unclear whether a physical network found can approximate all the infection networks of different diseases, which further limits the applicative value of existed methods. As a result, the methods that can achieve fine-grained intra-city epidemic propagation analysis and do not require detailed residential mobility empirical data are still highly desired.

In this paper, we employ a two-step method for metapopulation based epidemic propagation analysis. Step I is to understand the propagation system by inferring the underlying disease infection network. A novel model called D$^2$PRI is proposed to reduce individual network inference into sub-population network inference, and the power-law distribution prior and data prior are also incorporated for enhancements. Step II is to predict the infection propagation by using a metapopulation SIR model that allows visitors transmission between any two sub-populations. We conduct experiments on a real-life clinical report data set about the airborne disease in the the famous Shenzhen city in southern China. The D$^2$PRI model and the metapopulation SIR model show more excellent performances than some baseline methods in various tasks such as network inference, infection prediction and outbreaks simulation. We also apply our method in real-world applications. 




\section{Modeling Infection Propagation in Metapopulations}
\label{sec:propagation_model}

In this section, we start from introducing the classic Susceptible-Infectious-Recovered (SIR) model for single population modeling, and then extend it to describe the propagation of epidemic in an intra-city metapopulation.

\subsection{The Single-Population SIR Model}
\label{sec:sir}

In this study, we adopt the classical SIR model to describe the dynamic process of epidemic propagation. Given a population that contains a group of individuals, the SIR model divides the individuals as three compartments (states): the $S$ states is for the susceptible individuals, $I$ for the infectious, and $R$ for the recovered.

The SIR model assumes that all individuals have the same probability to contact each other. For a population with $P$ individuals, we use $s{(t)}$, $i{(t)}$, $r{(t)}$ to denote the numbers of individuals in the three states at time $t$. Therefore, given a {\em contact probability} $\alpha_1$, there are total $\alpha_1 \cdot { s{(t)}i{(t)}}$ times of contacts between the susceptible and infectious in the unit time $t$. Assuming the {\em infection probability} of a contact is $\alpha_2$, the number of susceptible individuals getting infected and switching to the $I$ state is $\alpha\cdot s(t)i(t)$, where $\alpha = \alpha_1 \cdot \alpha_2$ is named as the {\em Infection Rate}. By further assuming a $\beta$  fraction of infectious individuals are cured during an unit time, the number of individuals switching from $I$ state to the $R$ state is $\beta\cdot i(t)$.

Given the above, $s{(t)}$, $i{(t)}$, $r{(t)}$ have the following dynamics~\cite{sir}:
\begin{equation}\label{equ:basic_SIS}
\begin{split}
\frac{\mathrm{d} s{(t)}}{\mathrm{d} t}\;&= - \alpha \cdot { s{(t)}i{(t)}}, \\
\frac{\mathrm{d} i{(t)}}{\mathrm{d} t}\;&= \alpha\cdot {s{(t)}i{(t)}} - \beta\cdot i(t), \\
\frac{\mathrm{d} r{(t)}}{\mathrm{d} t}\;&= \beta\cdot i(t),
\end{split}
\end{equation}
which implies that $s{(t)}+i{(t)}+r{(t)} = P,~\forall~t$.

\subsection{The Metapopulation SIR Model}
\label{sec:sir_meta}

The basic SIR model implicitly assumes a homogeneous infection network between individuals and thus can only model epidemic propagation in a single population. Here, we extend the SIR model to the metapopulation scenario.

A {\it metapopulation} refers a group of separated sub-populations of the same species which interact at some level. Given a metapopulation with $N$ sub-populations, we denote the total number of individuals in sub-population $n$ as $P_n$, and the numbers of individuals in the $S,I,R$ states at time $t$ as $s_n{(t)}$, $i_n{(t)}$, $r_n{(t)}$, respectively. Between two sub-populations $n$ and $m$, the interaction strength is defined as $h_{nm}$, which is the average volume of visitors from $n$ to $m$ in a unit time. Given the above, the dynamic relationship of $s_n{(t)}$, $i_n{(t)}$, $r_n{(t)}$ is expressed as
\begin{equation}\label{equ:network_SIS}\small
\begin{split}
\frac{\mathrm{d} s_{n}{(t)}}{\mathrm{d} t} &= - \alpha\cdot s_n{(t)}\sum_{m=1}^N \left(\frac{h_{mn}}{P_m}+\frac{h_{nm}}{P_n}\right)i_{m}{(t)}, \\
\frac{\mathrm{d} i_{n}{(t)}}{\mathrm{d} t} &= \alpha\cdot s_n{(t)}\sum_{m=1}^N \left(\frac{h_{mn}}{P_m}+\frac{h_{nm}}{P_n}\right)i_{m}{(t)} - \beta\cdot i_{n}{(t)},\\
\frac{\mathrm{d} r_{n}{(t)}}{\mathrm{d} t} &= \beta\cdot i_{n}{(t)}.
\end{split}
\end{equation}

We here give detailed explanations to the first two equations in Eq.~\eqref{equ:network_SIS}. In a metapopulation, a susceptible individual of sub-population $n$ may contact with infectious individuals from three sources: $i$) The infectious in the same sub-population with a total number of $i_n(t)$, which will result in $\alpha\cdot s_n(t) i_n(t)$ new infectious in $n$, where $\alpha$ is the infection rate; $ii$) The infectious visitors from other sub-populations. The probability for an individual in $m$ visiting $n$ can be estimated by $h_{mn}/P_m$, so the new infectious in $n$ totals $\alpha\cdot s_n(t) \sum_{m\neq n} (h_{mn}/P_m) i_{m}(t)$; $iii)$ The infectious of other sub-populations who are contacted by the susceptible visitors from $n$. The probability of an individual in $n$ visiting $m$ can be estimated by $h_{nm}/P_n$, so the resulted new infectious in $n$ is $\sum_{m\neq n} \alpha\cdot s_n(t)(h_{nm}/P_n) i_{m}{(t)}$. For convenience, we define $h_{nn} = P_n/2$, so the total number of new infections caused by the three types of contacts is: $\alpha\cdot s_n{(t)}\sum_{m=1}^N \left(h_{mn}/P_m+h_{nm}/P_n\right)i_{m}{(t)}$.

Eq.~\eqref{equ:network_SIS} models epidemic propagation in a metapopulation as a dynamic change of individual numbers in different states. Given the initial states  $s_n{(0)}$, $i_n{(0)}$, $r_n{(0)}$ and the infection and recovery rates $\alpha$ and $\beta$ empirically, we can use Eq.~\eqref{equ:network_SIS} to recursively predict the epidemic propagation process in a metapopulation.

\subsection{Problem Formulation}
When applying Eq.~\eqref{equ:network_SIS} for real-life epidemic propagation prediction in a metapopuation, we still face a serious problem: How to set the individual mobility volumes $h_{nm}$, $\forall~n,m$? This is not a trivial issue, since $h_{nm}$'s are often unobservable and are different from city to city. Although there exist some studies in the literature that claimed to find some physical networks like cell-phone user mobility network~\cite{malaria} that can explain the propagation of some disease, the generality and availability of these physical networks are very limited for different types of infectious diseases and different application scenarios.

In this study, we attempt to solve the above problem from a very different perspective. That is, if we can collect the time series data about the number of infected people in a metapopulation, we can infer the dynamics of the propagation system behind the infection data, and $h_{nm}$'s can be regarded as the key parameters of the system and can be inferred accordingly. Following this idea, the problem of modeling epidemic propagation in a metapopulation can be decomposed into two steps. Step I is to {\bf understand} the propagation system for a specific infectious disease by inferring its parameters, and Step II is to use the system (Eq.~\eqref{equ:network_SIS}) to {\bf predict} the future propagation for epidemic control and prevention.

It is obvious that Step I is the key for solving the whole problem, so we focus on understanding the propagation system in the following Sect.~3 and Sect.~4. Specifically, we view a sub-population of a metapopulation as a node, and the individuals' visits between two sub-populations as directed edges. So the epidemic propagation system can be viewed as a directed network, with $h_{mn}$'s being the network parameters to be inferred.

\textbf{\textit{Remark.}}~Transforming Step I into a network inference problem has three obvious advantages. The first is to set the parameters in Eq.~\eqref{equ:network_SIS} more accurately in an objective way. The second is to enhance the generality of the whole solution to fit different infectious diseases --- we can learn different parameters for distinct diseases. The third is to help us to gain deep insight into the epidemic propagation system, which is crucial for making proper decisions for disease control and prevention. We will revisit the last point in the real-world application section below.

\section{Network Inference Model}
\label{sec:network_model}

In this section, we formalize the dynamic relationship defined in Eq.~\eqref{equ:network_SIS} as a network interaction model, and propose a network inference framework to implicitly infer the individual mobility volume $h_{nm}$.

\subsection{Network Interaction Model}
\label{sec:network_interaction}

We discretize the time line as a sequence of time slices, {\it i.e.}, $t = \{1, 2, \cdots, T\}$, and assume $s_n{(t)}$, $i_n{(t)}$, $r_n{(t)}$ of a sub-population are invariable in a time slice. We define $\delta_n{(t)}$ as the number of individuals newly infected in the time slice $t$, {\it i.e.}, the number of individuals switching from $S$ to $I$ during $t$ to $t+1$. According to the dynamic relations defined in Eq.~\eqref{equ:network_SIS}, $\delta_n{(t)}$ is calculated as
\begin{equation}\label{equ:prediction}
\begin{split}
\delta_n{(t)} = - \int_t^{t+1} \frac{\mathrm{d} s_{n}{(x)}}{\mathrm{d} x} \mathrm{d} x = \alpha s_n{(t)}\sum_{m=1}^N \left(\frac{h_{mn}}{P_m}+\frac{h_{nm}}{P_n}\right)i_{m}{(t)}.
\end{split}
\end{equation}

We model the metapopulation as a network with $N$ nodes and ${N\times N}$ edges connecting the nodes. The nodes indicate sub-populations and the edges indicate interactions between sub-populations. For the node $n$, we define a state variable ${u}_n^{(t)}$ to describe the current condition of the node $n$ at time $t$ as follows:
\begin{equation}\label{equ:epi_uv}
u_n^{(t)} = \frac{\delta_n{(t)}}{s_n{(t)}}.
\end{equation}
In the epidemiology, $u_n^{(t)}$ is called the {\it Incidence Rate} of a sub-population, which refers the number of new cases per population at risk (susceptible) in a given time period~\footnote{\url{https://en.wikipedia.org/wiki/Incidence_(epidemiology)}}. $u_n^{(t)}$ is an important variable in the epidemic propagation. $s_n{(t)}, i_n{(t)}, r_n{(t)}$ of a sub-population for any given time $T$ can all use the historical incidence rates $\mathbf{u}_n^{(<T)} = \{{u}_n^{(1)}, {u}_n^{(2)}, \ldots, {u}_n^{(T-1)}\}$ to calculate:
\begin{equation}\label{equ:epi_sir}
\begin{split}
s_n(T) =& f_s\left(\mathbf{u}_n^{(<T)}\right) = P_n \prod_{t=1}^{T-1} \left(1-u_n^{(t)}\right),\\
i_n(T) =& f_i\left(\mathbf{u}_n^{(<T)}\right) = \sum_{t=1}^{T-1} (1-\beta)^{t-T} \delta_n{(T)},\\
= &\sum_{t=1}^{T-1} (1-\beta)^{t-T} u_n^{(T)}{P_n} \prod_{t=1}^{T-1} \left(1-u_n^{(t)}\right),\\
r_n(T) =& P - s_n(T) - i_n(T) .
\end{split}
\end{equation}

For the edge from the node $n$ to $m$, we define its weight $g_{nm}$ as
\begin{equation}\label{equ:epi_network}
g_{nm} := \alpha\left(\frac{h_{mn}}{P_m}+\frac{h_{nm}}{P_n}\right),~\forall~n,m.
\end{equation}
It is easy to see that the physical meaning of $g_{nm}$ is the two-way mobility intensity between two sub-populations multiplied by the infection rate $\alpha$. Denote the matrix $\mathbf{G} \in \mathbb{R}^{N \times N}$ with the elements $g_{nm}$ as the network adjacent matrix. We call the network $\mathbf{G}$ as the {\em Infection Network}. It is obvious that $\mathbf{G}$ is a symmetric matrix, although the whole network is directed. Further let ${v}_n^{(t)} = i_n{(t)}$. By inserting Eq.~\eqref{equ:epi_uv} and Eq.~\eqref{equ:epi_network} into the Eq.~\eqref{equ:prediction}, we have
\begin{equation}\label{eq:interaction}
u_n^{(t)} = \sum_{i=1}^N v_m^{(t)} g_{mn},~\forall~n.
\end{equation}

\textbf{\textit{Remark.}}~Note that from Eq.~\eqref{equ:epi_sir}, ${v}_n^{(t)} = i_n{(t)}$ and $s_n{(t)}, r_n{(t)}$ can all be calculated using $\mathbf{u}_n^{(<T)}$. Therefore, if the matrix $\mathbf{G}$ is available, we can use Eq.~\eqref{eq:interaction} as a ``condensed'' yet equivalent system for epidemic propagation prediction. In other words, by taking a network perspective to a metapopulation and introducing new states $u$ and $v$, our problem reduces to the inference of $\mathbf{G}$ (rather than more detailed $h_{mn}$'s in Eq.~\eqref{equ:network_SIS}). We describe it formally below.

\subsection{Network Inference Problem}
\label{sec:basic_model}


We denote the states $u,v$ of all sub-populations at time $t$ as $\mathbf{u}^{(t)} =$ $(u_1^{(t)},$ $\ldots,$ $u_n^{(t)},$ $\ldots,$ $u_N^{(t)})^\top$ and $\mathbf{v}^{(t)} =(v_1^{(t)},$ $\ldots,$ $v_n^{(t)},$ $\ldots,v_N^{(t)})^\top$.
The interactions of sub-populations over the infection network are expressed as
\begin{equation}\label{equ:network_spreading}
\mathbf{u}^{(t)} = \mathbf{G}\mathbf{v}^{(t)} + \mathbf{e}^{(t)},
\end{equation}
where $\mathbf{e}^{(t)} = ({e}_1^{(t)}, {e}_2^{(t)}, \ldots, {e}_N^{(t)})^\top$ is introduced to model random noises in empirical data. Then, the {\em Network Inference problem} of the network interaction model in Eqs.~\eqref{equ:epi_uv} - \eqref{eq:interaction} is defined as:

{\em Definition 1: Network Inference Problem.} Given observable states series $\mathbf{U} = \{\mathbf{u}^{(1)},$ $\mathbf{u}^{(2)},$ $\ldots,$ $\mathbf{u}^{(T)}\}$ and $\mathbf{V} = \{\mathbf{v}^{(1)},$ $\mathbf{v}^{(2)},$ $\ldots,$ $\mathbf{v}^{(T)}\}$ of a metapopulation propagation network, inferring the adjacent matrix $\mathbf{G}$ according to Eq.~\eqref{equ:network_spreading}.\hfill$\blacksquare$

In practical, due to the data availability issue, we use the number of newly infected individuals in a unit time, {\it i.e.}, $\delta_n^{(t)}$, to calculate $u_n^{(t)}$ and $v_n^{(t)}$ as follows:
\begin{equation}\label{}\small
\begin{split}
u_n^{(T)} = \frac{\delta_n^{(T)}}{P_n - \sum_{t=1}^{T-1} \delta_n^{(t)}},~v_n^{(T)} = \sum_{t=1}^{T-1} (1-\beta)^{T-t-1} \delta_n^{(t)}.
\end{split}
\end{equation}
Compared with other variables, $\delta_n^{(t)}$ is easier to obtain, for example, from daily clinic reports of CDC. The recovery rate $\beta$ can be set as follows. For diseases that require hospitalization, $\beta$ can be calculated according to the number of hospitalizations; otherwise, we assume an infectious individual will recover after a given time period based on the actual situation.

\subsection{The Basic Network Inference Model}

We assume the noise ${e}_n^{(t)}$ in Eq.~\eqref{equ:network_spreading} is an {\it i.i.d.} random variable that follows a zero-mean Gaussian distribution, {\it i.e.}, ${e}_n^{(t)} \thicksim\mathcal{N}(0,\sigma^2_e),~\forall~n,t$. Given the network state ${\mathbf{v}}^{(t)}$ and the interaction network $\mathbf{G}$, the conditional probability distribution of $\mathbf{u}^{(t)}$ is calculated as
\begin{equation}\label{equ:likelihood}
P\left(\mathbf{u}^{(t)} \middle| {\mathbf{G}}, {\mathbf{v}}^{(t)}\right) = \prod_{n=1}^N \mathcal{N}\left({u}_n^{(t)}\middle |{\mathbf{g}}_{n:} \cdot {\mathbf{v}}^{(t)}\right),
\end{equation}
where ${\mathbf{g}}_{n:}$ is the $n$-th row vector of ${\mathbf{G}}$. Then the log Likelihood probability of $\mathbf{u}^{(t)}$ is formulated as
\begin{equation}\label{}
\log P\left(\mathbf{u}^{(t)} \middle| {\mathbf{G}}, {\mathbf{v}}^{(t)}\right) \propto -\frac{1}{\sigma^2_e} \sum_{n=1}^N \left( u_n^{(t)} - {\mathbf{g}}_{n:} \cdot {\mathbf{v}}^{(t)} \right)^2.
\end{equation}
Therefore, the {\em Maximum Likelihood Estimation} (MLE) of $\mathbf{G}$ for $T$ interaction rounds is to minimize the objective function
\begin{equation}\label{equ:obj_basic}
\mathcal{J}_1 = \frac{1}{\sigma^2_e}  \sum_{t=1}^{T} \left\| \mathbf{u}^{(t)} - {\mathbf{G}}\cdot {\mathbf{v}}^{(t)} \right\|^2_2.
\end{equation}

\section{Incorporating Priori Knowledge}
\label{sec:Priori}

In this section, we propose an improved network inference model by incorporating two types of priors: the power-law distribution prior and the data prior.

\subsection{Power-Law Distribution Priori}



The first type of priori is the priori distribution of network edge weights in $\mathbf{G}$. Traditional methods usually use the Gaussian (L2 regularization) or Laplace (L1 regularization) distributions as priori distributions of variables to be inferred~\cite{prml}. However, the Gaussian and Laplace distributions are not suitable for our model. As reported in many empirical studies~\cite{spatial_network}, spatial individual mobility networks usually behave as scale-free networks --- the degree of network nodes follows a power-law distribution rather than Gaussian or Laplace distribution. Therefore, we need to incorporate the power-law prior to regularize the node degrees in $\mathbf{G}$.

We assume the out-degree of node $n$ in $\mathbf{G}$ follows a power-law distribution, which means
\begin{equation}\label{}
P\left(\sum_{m=1}^N {g}_{nm} = x \right) = a \cdot x^{-k},
\end{equation}
where $k$ is usually set as $2<k<3$. For the interaction network, the priori probability of $\mathbf{G}$ is
\begin{equation}\label{eq:pl_priori_probability}
P\left(\mathbf{{G}}\right) = \prod_{n=1}^{N} a\cdot \left(\sum_{m=1}^N {g}_{nm}\right)^{-k}.
\end{equation}
Because the $\mathbf{G}$ is a symmetrical matrix, our model only considers the out-degree.

Inserting the priori probability into the likelihood probability in Eq~\eqref{equ:likelihood}, we obtain the posterior distribution of ${\mathbf{G}}$ for given ${\mathbf{v}}^{(t)}$ and $\mathbf{u}^{(t)}$ as follows:
\begin{equation}\label{}
P\left({\mathbf{G}} \middle|\mathbf{u}^{(t)}, {\mathbf{v}}^{(t)}\right) = \frac{ P\left(\mathbf{u}^{(t)} \middle| {\mathbf{G}}, {\mathbf{v}}^{(t)}\right) P\left({\mathbf{G}} \right)}{P\left(\mathbf{u}^{(t)}\right)}.
\end{equation}
Then the log posterior distribution of ${\mathbf{G}}$ is
\begin{equation}\label{}\small
\begin{split}
\ln P&\left({\mathbf{G}} |\mathbf{u}^{(t)}, {\mathbf{v}}^{(t)}\right) \\
\propto & -\frac{1}{\sigma^2_e} \sum_{n=1}^N \left( u_n^{(t)} - {\mathbf{g}}_{n:} \cdot {\mathbf{v}}^{(t)} \right)^2 - k \sum_{n=1}^N \ln \left(\sum_{m=1}^N {g}_{nm}\right).
\end{split}
\end{equation}
Therefore, the Maximum A Posteriori (MAP) estimation of ${\mathbf{G}}$ is to minimize the objective function $\mathcal{J}_2$ as follows:
\begin{equation}\label{equ:obj_pl}
\mathcal{J}_2 = \sum_{t=1}^{T} \left\| \mathbf{u}^{(t)} - {\mathbf{G}}\cdot {\mathbf{v}}^{(t)} \right\|^2_2 + \lambda \sum_{n=1}^N \ln \left(\sum_{m=1}^N {g}_{nm}\right).
\end{equation}
where $\lambda = k\sigma_e^2$ is a preset parameter.

\subsection{Data Priori}

The other type of priori to be incorporated is the knowledge extracted from related data. In our model, the network edge weight $g_{nm}$ is proportional to the individual mobility intensity between the sub-population $n$ and $m$. Therefore, we could use some mobility related data to estimate $g_{nm}$. For example, if $g_{nm}$ denotes resident visiting between two urban zones, taxi GPS trajectory, bus/metro smart card records, or LBS check-in data could be considered as priori knowledge. In our model, we adopt a linear regression-based regularization method to incorporate the date priori.

Suppose altogether we have $K$ features extracted from related data sets. Then for any ${g}_{nm}\in{\mathbf{G}}$, we have a feature vector $\mathbf{x}_{nm} = ( x_{nm,1},$ $\ldots,$ $x_{nm,k},$ $\ldots,$ $x_{nm,K})^{\top}$, where $x_{nm,k}$ is the $k$-th feature. Then, a linear regression is used to model the relations between $g_{nm}$ and $\mathbf{x}_{nm}$ as
\begin{equation}\label{eq:linear_regression}
{g}_{nm} = \mathbf{w}^{\top}\mathbf{x}_{nm} + e_{nm},
\end{equation}
where $\mathbf{w} = \left(w_1, \ldots, w_k, \ldots, w_{K-1} \right)^{\top}$ is a trainable weight vector, and $e_{nm}$ is an {\it i.i.d.} random regression error.

We define a tensor $\mathcal{X}\in \mathbb{R}^{N\times N \times K}$ composed by $\mathbf{x}_{nm}$ as the ($n,m$) fiber. The linear regression in Eq.~\eqref{eq:linear_regression} can be written in a matrix form as
\begin{equation}\label{}
{\mathbf{G}} = \mathcal{X} \times_k \mathbf{w} + \mathbf{E},
\end{equation}
where $\times_k$ is the $k$-mode product~\cite{tensor} between tensor $\mathcal{X}$ and vector $\mathbf{w}$, and $\mathbf{E}$ is a matrix composed by $e_{nm}$.

We adopt a zero-mean Gaussian noise with variance $\sigma^2_{x}$ to model the regression error as $e_{nm}\thicksim \mathcal{N}(0,\sigma^2_{x})$. Then the conditional distribution of ${\mathbf{G}}$ with a regression model determined by $\mathbf{w}$ is given by
\begin{equation}\label{}
P\left({\mathbf{G}}\middle|\mathbf{w},\mathbf{x}\right) = \prod_{m=1}^N \prod_{n=1}^{N} \mathcal{N}\left({g}_{nm} \middle| \mathbf{w}^{\top} \mathbf{x}_{nm}, \sigma^2_{x}\right).
\end{equation}
We then introduce a zero-mean Gaussian prior on the regression weight vector $\mathbf{w}$, which gives
\begin{equation}\label{}
P\left(\mathbf{w}\right) = \prod_{k=1}^K \mathcal{N}\left(w_k|0, \sigma_{w}^2\right).
\end{equation}
The log posterior probability distribution of the regression weight vector $\mathbf{w}$ and network adjacent matrix ${\mathbf{G}}$ is
\begin{equation}\label{}
\begin{aligned}
\ln  P&\left({\mathbf{G}},\mathbf{w}|\mathbf{x}\right) = \ln{P\left({\mathbf{G}}\middle|\mathbf{w},\mathbf{x}\right) P\left(\mathbf{w}\right)} \\
\propto & -\frac{1}{\sigma^2_{x}} \sum_{n=1}^N \sum_{m=1}^{N} \left({g}_{nm} - \mathbf{w}^\top \mathbf{x}_{nm}\right)^2 -\frac{1}{\sigma^2_{w}} \sum_{k=1}^K w_k^2.
\end{aligned}
\end{equation}
Therefore, maximizing posterior probability of $\mathbf{w}$ and ${\mathbf{G}}$ for given data priori $x$ is equivalent to minimizing the objective function $\mathcal{J}_3$ as
\begin{equation}\label{equ:obj_lr}
\mathcal{J}_3 = \frac{1}{\sigma^2_{x}} \left \| {\mathbf{G}} - \mathcal{X} \times_k \mathbf{w} \right \|^2_F + \frac{1}{\sigma^2_{w}} \left \|\mathbf{w}\right \|_2^2,
\end{equation}
where $\|\cdot\|_F$ is the Frobenius Norm.

\subsection{The D$^2$PRI Model}

We here integrate the objective functions $\mathcal{J}_2$ and $\mathcal{J}_3$ to get a joint model, which is named as D$^2$PRI (power-law Degree and Data Priori jointly Regularized non-negative network Inference). The objective function of D$^2$PRI is
\begin{equation}\label{equ:final_obj}
\begin{aligned}
\mathop{\arg\min}_{{\mathbf{G}},\mathbf{w}}  \mathcal{J} = &\sum_{t=1}^{T} \left\| \mathbf{u}^{(t)} - {\mathbf{G}}\cdot {\mathbf{v}}^{(t)} \right\|^2_2 + \lambda \sum_{n=1}^N \ln \left(\sum_{m\neq n} {g}_{nm}\right) \\
&+ \eta \left \| {\mathbf{G}} - \mathcal{X} \times_k \mathbf{w} \right \|^2_F +  \mu \left \|\mathbf{w}\right \|_2^2\\
s.t. &\;\;{\mathbf{G}} \geq 0, \mathbf{w} \geq 0,
\end{aligned}
\end{equation}
where $\eta  = \sigma^2_{e}/\sigma^2_{x}$, $\mu = \sigma^2_{e}/\sigma^2_{w}$ and $\lambda = k\sigma_e^2$ are preset parameters. Note that since the individual mobility intensity cannot be negative we introduce a non-negativity constraint to $\mathbf{G}$. Moreover, we also introduce a non-negativity constraint of $\mathbf{w}$ to reduce solution space. It requires the features $\mathbf{x}_{nm}$ to have positive correlations with the individual mobility intensity, which is easy to be satisfied in data preprocessing.

\section{Optimization}
\label{sec:optimization}

In this section, we propose a Semi-supervised Proximal Gradient Descent (SPGD) algorithm to solve the D$^2$PRI model.

As shown in Algorithm~\ref{alg:algorithm}, SPGD iteratively optimizes $\mathcal{J}$ defined in Eq.~\eqref{equ:final_obj}. In each iteration, the algorithm alternately uses $\mathbf{G}$ to train $\mathbf{w}$ and uses $\mathbf{w}$ to predict $\mathbf{G}$, which could be considered as a semi-supervised training process for a model to predict $\mathbf{G}$. Specifically, in the $l$-th iteration, we use the Proximal Gradient Descent to update $\mathbf{G}_{l}$ from $\mathbf{G}_{l-1}$ with $\mathbf{w}_{l-1}$ as
\begin{equation}\label{eq:update_g3}
\begin{aligned}
{\mathbf{G}}_{(l)} = \max\left(0, {\mathbf{G}}_{(l-1)} - \frac{1}{L}\frac{\partial\mathcal{J}\left(\mathbf{G}_{(l-1)}|\mathbf{w}_{(l-1)}\right)}{\partial{\mathbf{G}}_{(l-1)}} \right),
\end{aligned}
\end{equation}
and train ${\mathbf{w}}_{(l)}$ using $\mathbf{G}_{l}$ as
\begin{equation}\label{eq:update_w3}
\begin{aligned}
{\mathbf{w}}_{(l)} = \max\left(0, {\mathbf{w}}_{(l-1)} - \frac{1}{L}\frac{\partial\mathcal{J}\left(\mathbf{w}_{(l-1)}|\mathbf{G}_{(l)}\right)}{\partial{\mathbf{w}}_{(l-1)}} \right).
\end{aligned}
\end{equation}
Here, $L$ is a Lipschitz constant that satisfies
$\left\| \frac{\partial\mathcal{J}}{\partial{\mathbf{Z}}_{1}} - \frac{\partial\mathcal{J}}{\partial{\mathbf{Z}}_{2}}\right \|_F\leq L $ $ \left\|  {\mathbf{Z}}_{1} - {\mathbf{Z}}_{2} \right\|_F, \forall~{\mathbf{Z}}_{1}, {\mathbf{Z}}_{2}$, where $\mathbf{Z}$ respectively represents $\mathbf{G}$ and $\mathbf{w}$ in~\eqref{eq:update_g3} and~\eqref{eq:update_w3}.

According to Eq.~\eqref{equ:final_obj}, the partial derivative of $\mathcal{J}$ to ${{g}}_{nm}$ and $w_{k}$ are calculated as
\begin{equation}\label{equ:Penalty}
\begin{aligned}
&\frac{\partial\mathcal{J}}{\partial {g}_{nm}}  &=& 2 \sum_{t=1}^{T} \left( {\mathbf{g}}_{n:} \cdot {\mathbf{v}}^{(t)} -  u_n^{(t)} \right) {v}_m^{(t)}  + \underbrace{\frac{\lambda}{\sum_{k =1}^{N} {g}_{nk}}}_{\mathrm{Penalty~Term}} \\
& & & + 2 \eta \left({g}_{nm} - \mathbf{w}^\top \mathbf{x}_{nm}\right)\\
&\frac{\partial\mathcal{J}}{\; \partial {w}_{k}\;} &=& 2 \eta \sum_{n=1}^N \sum_{m=1}^{N} \left({g}_{nm} - \mathbf{w}^\top \mathbf{x}_{nm}\right){x}_{nm,k} + 2 \mu w_k,\\
\end{aligned}
\end{equation}

\begin{algorithm}[t]
	\caption{Semi-supervised Proximal Gradient Descent (SPGD)}
	\begin{algorithmic}[1]\label{alg:algorithm}
		\REQUIRE $\left\{\mathbf{u}^{(t)}, {\mathbf{v}}^{(t)}, t\in \{1,2,\ldots, T\}\right\}$, $\lambda, \eta, \mu$
		\STATE {Initialization}: Randomize ${\mathbf{G}}_{(0)}$ and $\mathbf{w}_{(0)}$
		\FOR {$l=1, 2, \ldots$}
		\STATE Update ${\mathbf{G}}_{(l)}$ by Eq.~\eqref{eq:update_g3}.
		\STATE Update $\mathbf{w}_{(l)}$ by Eq.~\eqref{eq:update_w3}.
		\IF {converged}
		\STATE Return $\left( {\mathbf{G}}_{(l)}, \mathbf{w}_{(l)}\right)$.
		\ENDIF
		\ENDFOR
	\end{algorithmic}
\end{algorithm}

\textbf{\textit{Remark.}}~As shown in Eq.~\eqref{equ:Penalty}, the power-law degree regularization introduces a penalty term to the partial derivative of $\mathcal{J}$ {\it w.r.t.} $g_{nm}$. The penalty term is inversely proportional to the out-degree of node $n$, {\it i.e.} $\sum_{k =1}^{N} {g}_{nk}$. Therefore, if node $n$ has a large degree, the algorithm gives small penalty to $g_{nm}$, and $g_{nm}$ thus tends to converge to a large value, and vice versa. This is consistent with the ``Matthew Effect'' in scale-free networks~\cite{ba} --- a node with large degree has higher possibility to connect other nodes.

\section{Experiments}
\label{sec:experiments}

\subsection{Data Description}

We use a real-life data set collected from Shenzhen~\footnote{\url{https://en.wikipedia.org/wiki/Shenzhen}}, a major city in southern China with a population over 11 million, to verify the proposed model D$^2$PRI. The variables used in our model include the sub-population size $P_n$, the sub-population states $u_n^{(t)}$ and $v_n^{(t)}$, the data prior features $\mathbf{x}_{nm}$, and the human mobility intensity network $\tilde{\mathbf{G}}$. All these variables are set using real world data as follows.

\begin{figure}
	\centering
	\includegraphics[width=0.8\columnwidth]{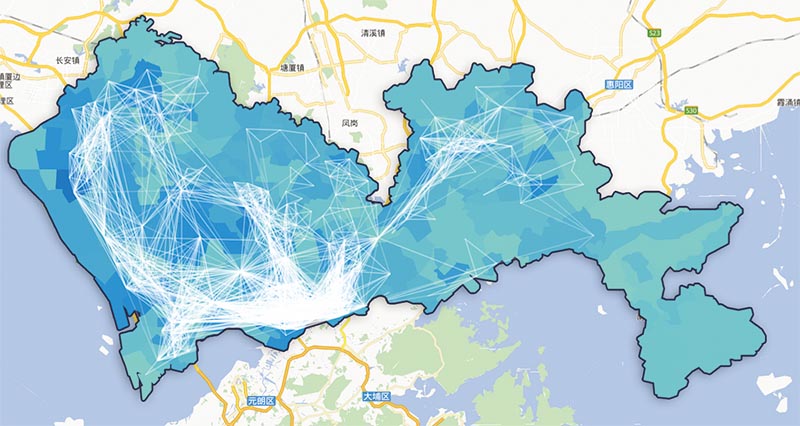}
	\caption{Zone segmentation of Shenzhen with human mobility intensity network.}\label{fig:map}
\end{figure}

We use the administrative boundaries to segment Shenzhen into 127 urban zones. The residents in the same zone are considered as a sub-population. The sub-population size $P_n$ is obtained from the population census data of Shenzhen. The map of these zones are plotted in Fig.~\ref{fig:map}, where the color denotes the population size of each zone, and the deeper the more.

\begin{figure}
	\centering
	\includegraphics[width=0.9\columnwidth]{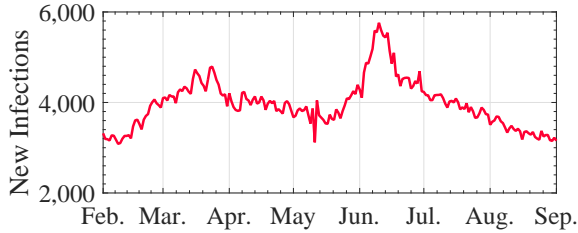}
	\caption{Daily new infections of airborne diseases in SZ.}\label{fig:flu_outbreak}
\end{figure}

The sub-population states $u_n^{(t)}$, $v_n^{(t)}$ are calculated from the clinical report data set offered by the Center for Disease Control and Prevention (CDC) of Shenzhen. The data set contains all airborne disease cases of Shenzhen from February to September in 2014. The fluctuation of daily new infection numbers in Shenzhen is plotted in Fig.~\ref{fig:flu_outbreak}. As can be seen, there are two outbreaks in the data, which happened in two periods, {\it i.e.}, March - May and May - August. In what follows, we call the two outbreaks as \textit{FirstOutbreak} and \textit{SecondOutbreak}, respectively, for convenience. The total infected persons in the two outbreaks respectively reached to 479 and 567 thousands.

In the experiments, we adopt two-feature $\mathbf{x}_{nm}$ as data priori. The first feature is extracted from a taxi trajectory data set, which contains the GPS trajectories of all taxies in Shenzhen during one week in April, 2014. We take the {\it traffic volumes} of taxies that carried passengers between two urban zones as a feature. The second feature is the {\it visitor volumes} estimated by the Gravity model~\cite{gravity}. The visitor volume between two zones $n,m$ is given by $x^g_{nm} = P_n\times P_m/D_{nm}^2$, where $D_{mn}$ is the distance between two zones.

$\tilde{\mathbf{G}}$ serves as a reference for the infection network $\mathbf{G}$, which is built using a mobile phone location data set containing the location (approximated by base station location) records for 11 million mobile phone users in Shenzhen during one week in April, 2013. The location of a user is updated in every half an hour. We count the number of visitors between urban zones as $h_{mn}$, and build a network with edge weights $\tilde{g}_{nm} = \left(\frac{h_{mn}}{P_m}+\frac{h_{nm}}{P_n}\right)$. Compared with the infection network ${g}_{nm}$ defined in Eq.~\eqref{equ:epi_network}, $\tilde{g}_{nm}$ does not contain the infection rate $\alpha$. Therefore, in our experiments, we use the similarity between $\tilde{g}_{nm}$ and ${g}_{nm}$ to measure model performance. Fig.~\ref{fig:map} plots the edges of $\tilde{\mathbf{G}}$ with top 10\% weights.

%


\subsection{Results of Network Inference}
\label{sec:quantitative}

The first experiment is network inference. In the experiment, we use the proposed model to infer the infection network form the state series $u_n^{(t)}$, $v_n^{(t)}$ of the \textit{FirstOutbreak}. The propagation of airborne diseases has close relations with resident mobilities. If the real propagation process of the airborne disease coincides with our model, the network inferred by our model ($\mathbf{G}$) should be very similar to the human mobility network extracted from the mobile phone data ($\tilde{\mathbf{G}}$). In the experiments, we use the {\it cosine similarity} between $\mathbf{G}$ and $\tilde{\mathbf{G}}$ as the measure of model performance. Our D$^2$PRI model is compared with the following baselines:  {\bf Basic}, which uses the objective function $\mathcal{J}_1$ in Eq.~\eqref{equ:obj_basic} with the non-negativity constraint of $\mathbf{G}$ to infer the network. {\bf PLPRI}, which uses the Basic model with power-law prior to infer the network. The objective function is $\mathcal{J}_2$ in Eq.~\eqref{equ:obj_pl} with the non-negativity constraint of $\mathbf{G}$. {\bf DatPRI}, which uses the Basic model with data priori to infer the network. The objective function is defined as $\mathcal{J}_1+\mathcal{J}_3$ with the non-negativity constraints of $\mathbf{G}$ and $\mathbf{w}$. {\bf L1PRI}, which uses the L1 term to regularize the Basic model. Its objective function is $mathcal{J}_4 = \sum_{t=1}^{T} \left\| \mathbf{u}^{(t)} - {\mathbf{G}}\cdot {\mathbf{v}}^{(t)} \right\|^2_2 + \zeta_1 \left\|{\mathbf{G}}\right\|_1$.  {\bf L2PRI}, which uses the L2 term to regularize the Basic model. Its objective function is $\mathcal{J}_5 = \sum_{t=1}^{T} \left\| \mathbf{u}^{(t)} - {\mathbf{G}}\cdot {\mathbf{v}}^{(t)} \right\|^2_2 + \zeta_2 \left\|{\mathbf{G}}\right\|_F^2$. The regularization parameters were set with trial and error.

\begin{figure}
	\centering
	\includegraphics[width=0.85\columnwidth]{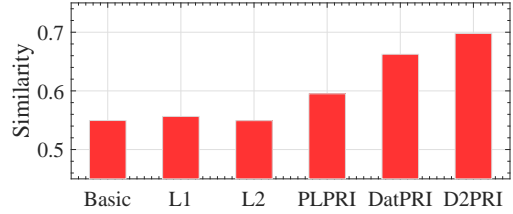}
	\vspace{-0.3cm}
	\caption{Comparison of network inference performances.}\label{fig:net_rec}
	\vspace{-0.4cm}
\end{figure}

Fig.~\ref{fig:net_rec} gives a comparison of the network inference performance between D$^2$PRI and the baselines. As shown in the figure, even the network inferred by the Basic model could achieve more than 0.5 similarity with the real mobility network. This implies that the proposed model framework can effectively describe the real-world disease propagation process. The performance of PLPRI is much better than L1PRI and L2PRI. The L1 and L2 regularizations actually did not achieve any significant performance improvement. This result demonstrates the merit of the power-law distribution prior in describing real-world human mobility patterns. The performance of DatPRI is better than PLPRI, which indicates the data prior can offer more accurate information than the distribution prior. Combining both data and power-law distribution priors, the proposed D$^2$PRI model achieved the best performance, which implies that D$^2$PRI coincides with the real-life airborne disease propagation process.

\subsection{Results of Infection Prediction}

The second experiment is infection prediction, in which we apply the network inferred in {\em FirstOutbreak} to predict the infections in {\em SecondOutbreak}.

As shown in Fig.~\ref{fig:flu_outbreak}, the two outbreaks appeared in succession, so the human mobility network should not have any dramatic change. Therefore, the experiment of applying the network of one outbreak for the prediction of the other outbreak can verify the robustness of the network inference model. Because the infection rate $\alpha$ in different outbreaks may change, we use the data in the first 1/3 days of {\em SecondOutbreak} to train an infection rate adjustment factor as $\tilde{\alpha} = \mathop{\arg\min}_{\tilde{\alpha}} \sum_{t=1}^{T} \left\| \mathbf{u}^{(t)} - \tilde{\alpha} \cdot \tilde{\mathbf{G}} {\mathbf{v}}^{(t)} \right\|^2_2$.

In the experiment, given any time point $T$ of {\em SecondOutbreak}, we use the $\mathbf{G}$ inferred in {\em FirstOutbreak} to iteratively predict $\delta_n^{(T+\Delta)}$, where $\Delta$ varies from one to seven days. The prediction performance is evaluated using the Mean Absolute Percentage Error (MAPE).

In addition to the baselines for the network inference experiment, we adopt two more time series models, {\it i.e.}, ARIMA~\cite{arima} and LSTM~\cite{lstm}, as the baselines. ARIMA is a benchmark of the classical time series prediction models, and LSTM represents deep learning methods. The ARIMA model treats the states of urban zones as time series to predict. The LSTM model uses $\mathbf{v}^{(t)}$ as features to predict $u_{n}^{(t)}$, and calculates $\delta_{n}^{(t)}$ using the method described in Sect.~\ref{sec:network_model}.

\begin{table}[t]
	\caption{Comparison of prediction performances.}
	\vspace{-0.2cm}
	\label{tab:prediction_performance}
	\centering  \footnotesize
	\resizebox{0.47\textwidth}{!}{
		\begin{tabular}{l|l|cccc }
			\toprule
			&{\bf Models} & {\bf 1-Day} & {\bf 3-Days} & {\bf 5-Days} & {\bf 7-Days}\\\midrule
			\multirow{7}{*}{MAPE}
			&{ D$^2$PRI} &{\bf  0.070} &  {\bf  0.190} & {\bf  0.300} & {\bf 0.409}  \\
			&{ DatPRI} & 0.072 & 0.194 &  0.306 &  0.418  \\
			&{ PLPRI}  & 0.074 & 0.201 &  0.319 &  0.436  \\
			&{ L1PRI}  & 0.076 & 0.207 &  0.328 &  0.450 \\
			&{ L2PRI}  & 0.076 & 0.206 &  0.327 &  0.449  \\
			&{ BASIC}  & 0.076 & 0.206 &  0.327 &  0.449 \\
			&{ ARIMA}  & 0.083 & 0.247 &  0.396 &  0.510 \\
			&{ LSTM}   & 0.073 & 0.200 &  0.310 &  0.422 \\
			\bottomrule
	\end{tabular}}
	\vspace{-0.3cm}
\end{table}

Table~\ref{tab:prediction_performance} lists the prediction performances of all models. As shown in the results, the D$^2$PRI model achieved the best performance than all baselines. The performance of PLPRI is better than L1PRI and L2PRI, and DatPRI is again better than PLPRI. These are consistent with the results of the network inference experiment. Even the Basic model has a better performance than ARIMA, which indicates that the infection network information is very important for epidemic prediction. The neural network based LSTM could model non-linear relations of daily infections among urban zones, so it achieved good performance. However, a weakness of neural network models is lacking of interpretability. In contrast, all variables in D$^2$PRI have clear physical meanings. We will show D$^2$PRI's interpretability advantage again in the application study below.

\subsection{Results of Outbreak Simulation}

The third experiment is epidemic outbreak simulation, in which we use the infection network inferred in {\em FirstOutbreak} to predict (simulate) all process of {\em SecondOutbreak}.
In the experiments, we use the first 10 days states of {\em SecondOutbreak} as an initial value to recursively calculate $S, I, R$ states in the rest of the outbreak. The infection rate $\alpha$ is also adjusted using the first 10 day's data. Compared with the short-time predictions, the long-term simulation is more valuable for epidemic control and prevention. The accurate long term disease propagation simulation can help the epidemic prevention personnel to prepare enough medical resources at the beginning stage of a outbreak. But meanwhile, the long-term simulation is also very challenging, because the simulation errors of each step can accumulate. If the infection network cannot model real condition very accurately, a minimal error or deviation may result in wide simulation divergence.

\begin{figure}[t]
	\centering
	\includegraphics[width=0.8\columnwidth]{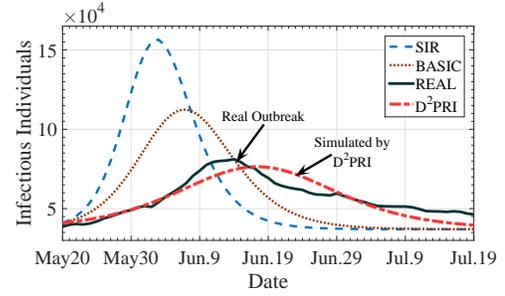}
	\caption{Comparison of epidemic outbreak simulations.}\label{fig:simulation}
\end{figure}

Fig.~\ref{fig:simulation} gives the simulation results, where the black line is the daily changes of real infectious individual numbers. The Basic and D$^2$PRI lines are simulated results using the network inferred by corresponding models. The SIR line is simulated by a non-networked SIR model, which considers all residents of Shenzhen as a single population.

As shown in Fig.~\ref{fig:simulation}, the non-networked SIR model obviously overestimated the outbreak speed and underestimated the duration. In the non-networked SIR model, all residents of the city have the same probability to contact others. A disease can rapidly propagate all over the population, which causes the outbreak to burst very quickly and soon disappear (most of individuals rapidly switch to the Recovered state). This implies that it is improper to assume all residents in the same city as a single population, although the assumption was adopted by many inter-city epidemic analysis works.

\begin{figure}[t]\centering
	\subfigure[Basic]{\label{fig:dgree_normal} \includegraphics[width=0.48\columnwidth]{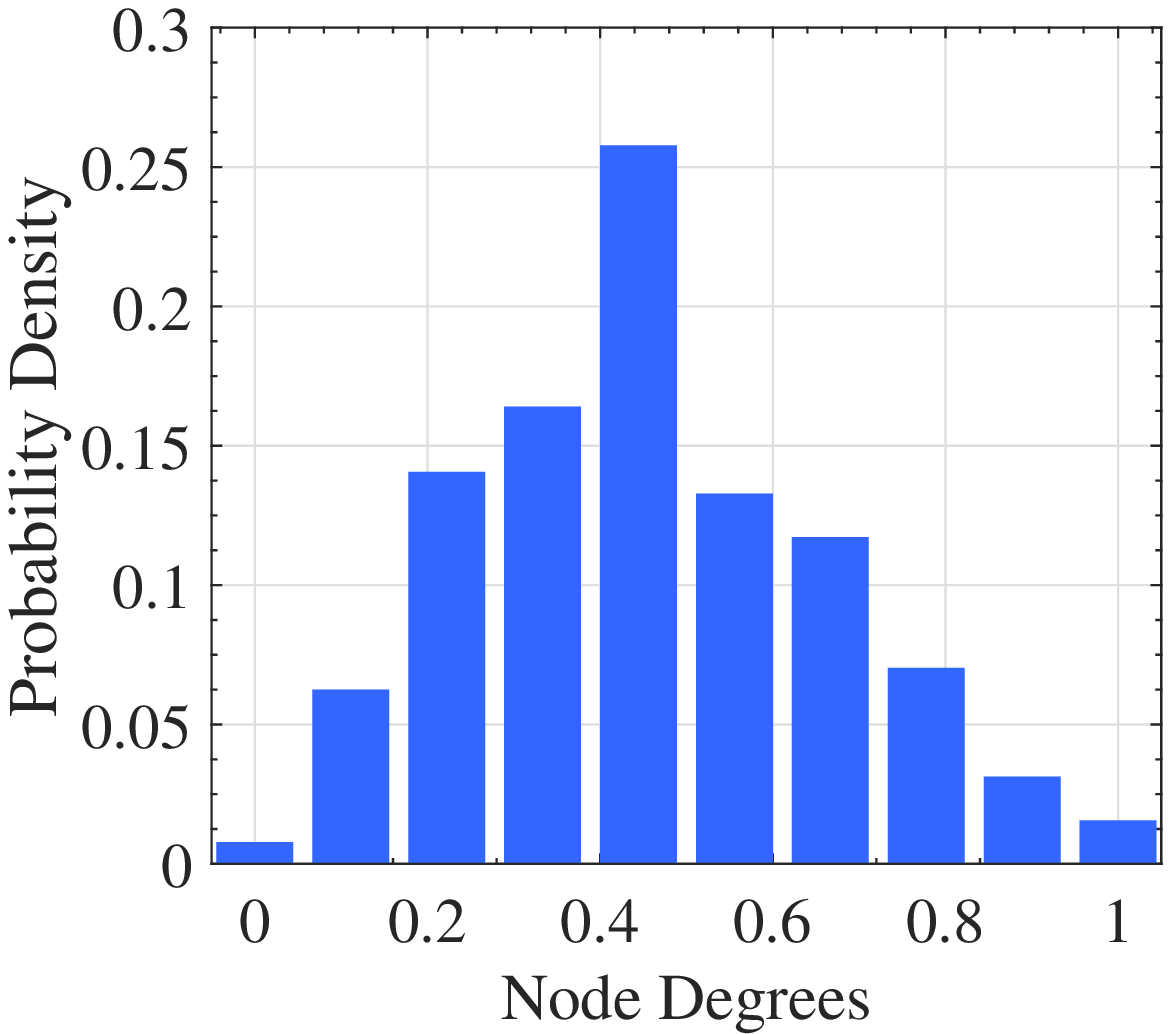}}
	\subfigure[D$^2$PRI]{\label{fig:dgree_pl} \includegraphics[width=0.48\columnwidth]{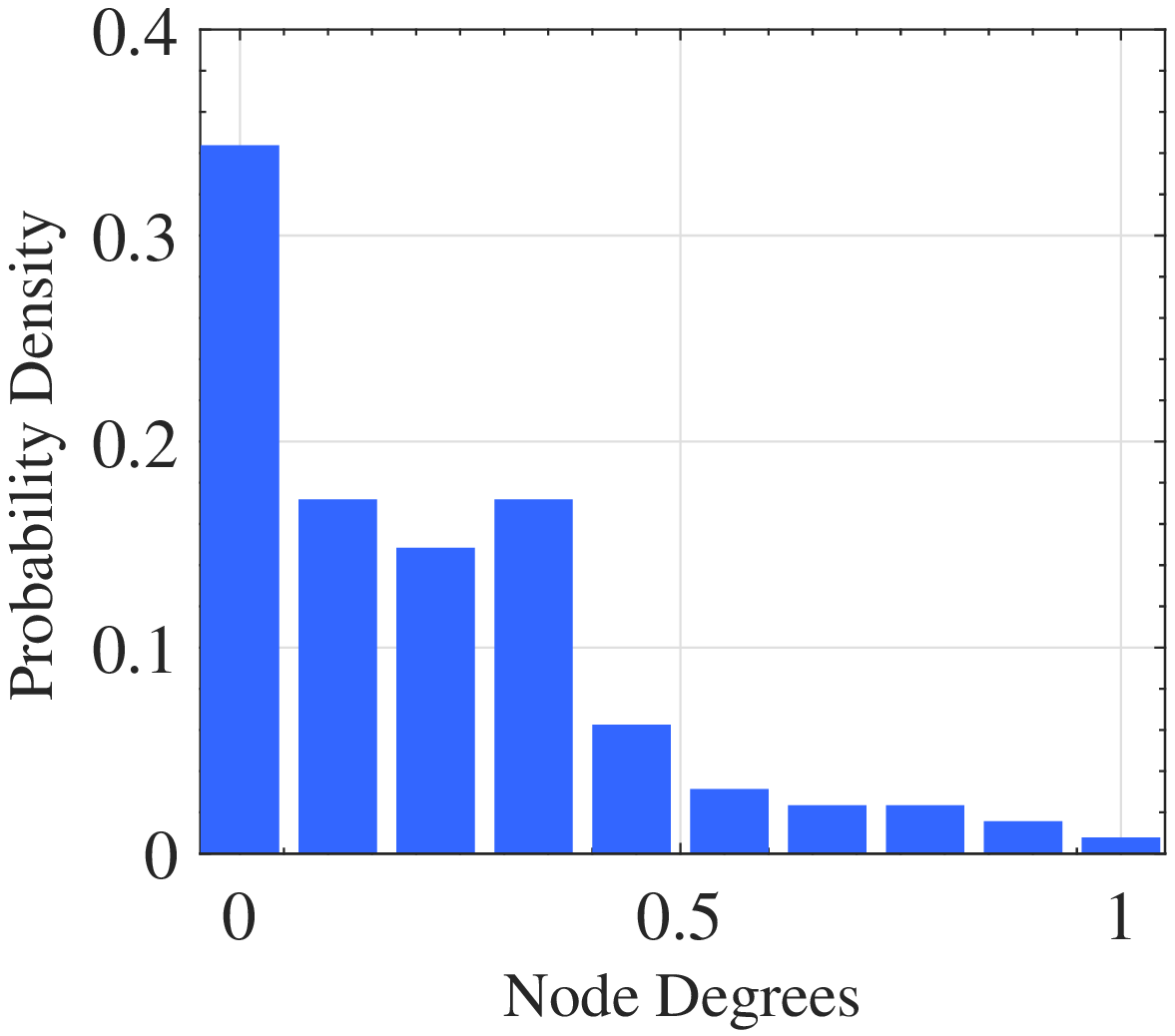}}
	\vspace{-0.3cm}
	\caption{Degree distributions of inferred networks.}
	\vspace{-0.4cm}
	\label{fig:dgree_disribution}
\end{figure}

The curve simulated by the Basic model has better performance than the non-networked SIR model. In the Basic model, except for the visitors, individuals can only contact with others within the same sub-population, which limits the outbreak speed of epidemics and increases the duration. Nevertheless, from the figure we can see the problem of overestimating breaking speed and intensity has not been fully eliminated in the Basic's curve. We seek the reason through analyzing the degree distribution of the Basic's network. The normalized degree distribution of the network is plotted in Fig.~\ref{fig:dgree_normal}, which tends to be a Poisson distribution and therefore the network is a Random Graph~\cite{newman2003structure}. Nodes in a Random Graph have homogeneous probability to connect with other nodes, which means that the residents in different sub-populations have the same cross-population contact probability. It does not coincide with the real world, where the cross-population contact probabilities for two neighboring zones and two remote zones are obviously different.

We also plot the degree distribution of the network inferred by D$^2$PRI in Fig.~\ref{fig:dgree_pl}. As shown in the figure, regularized by both the power-law distribution prior and data prior, the network degree distribution is much closer to a power-law distribution, which implies that the infection network is a scale-free network. A disease cannot propagate very quickly in a scale-free network due to the limitation of low degree nodes, but can continue for very long time because hub nodes with large degrees continually transmit disease from one sub-population to another. As shown in Fig.~\ref{fig:simulation}, by leveraging the power-law distribution and data priors, the proposed D$^2$PRI model simulates the outbreak process very accurately.


\textbf{\textit{Remark.}}~In the prediction and simulation experiments, the network inferred in one outbreak is used in the application of the other outbreak, which implies that the proposed model is very robust in different epidemic propagation conditions, and the inferred infection network is stable and universal for the Shenzhen city.

\section{Real-World Application}

The infection network inferred by our D$^2$PRI model has been applied by Shenzhen to detect important urban zones in epidemic propagation.

The traditional method directly uses the total infection number, {\it i.e.}, $\sum_{t=1}^T \delta_n^{(t)}$, as the importance measurement of urban zones. The implicit assumption behind it is: urban zones with more total infections are more important to the epidemic control work of CDC.

However, this straightforward method does not consider the impact of human mobility to disease transmission. Usually, urban zones with large population sizes have more infection numbers. Fig.~\ref{fig:infection_number} gives the map of importance for the urban zones in Shenzhen using the infectious number based method. The points on the map are geographic centers of urban zones, and the sizes of points denote importance level of each zone. As shown in the figure, the two most ``important'' zones locate at the downtown areas of Shenzhen, where the population densities are relatively higher. After excluding the two zones, however, the importance of the remaining zones seem are very similar to each other. This type of zone importance cannot give adequate help to epidemic control and prevention.

\textbf{\begin{figure}[t]\centering
		\subfigure[Importance by Infection Numbers]{\label{fig:infection_number} \includegraphics[width=0.85\columnwidth]{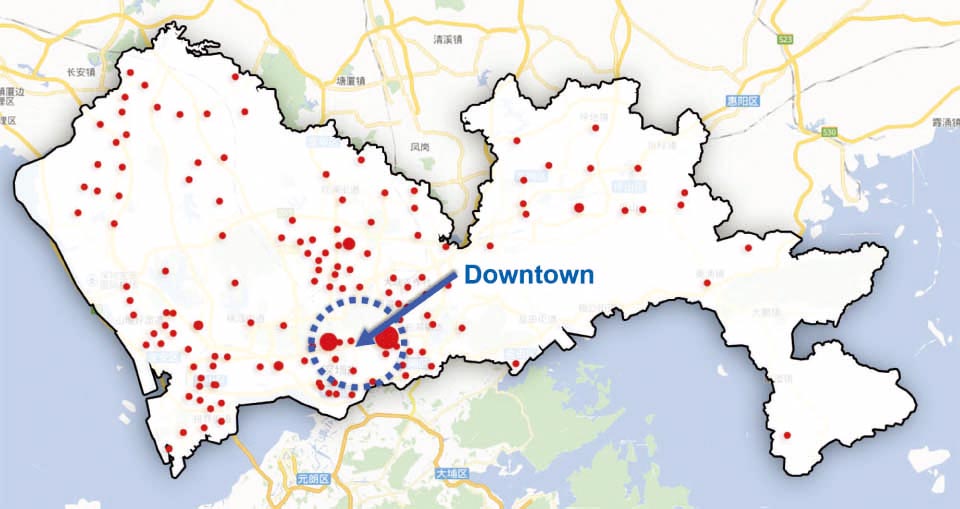}}
		\subfigure[Importance by Pagerank Scores from the D$^2$PRI Network]{\label{fig:pagerank} \includegraphics[width=0.85\columnwidth]{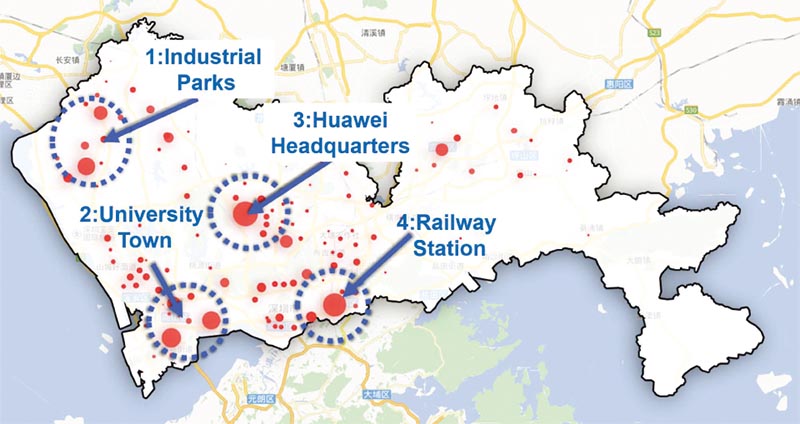}}
		\vspace{-0.3cm}
		\caption{Comparison of importance evaluation methods.}
		\vspace{-0.5cm}
		\label{fig:pagerank_vs_infection}
	\end{figure}
}

In our method, we applied a PageRank algorithm on the infection network inferred by the D$^2$PRI model in {\em FirstOutbreak}. The ranking scores were used as the importance measurement. Fig.~\ref{fig:pagerank} gives a map of the pagerank importance. As shown in the figure, the high score zones are geographically clustered on four areas: The first is in the northwest area of Shenzhen, which is a gathering place of {\it industrial parks}. The second is in the southwest area, which is the {\it university town}. The third is in the central region of Shenzhen, which is the {\it headquarter of the Huawei company}, the biggest high-tech enterprise in Shenzhen with more than 0.1 million employees working in the headquarter. The fourth is the {\it Shenzhen Railway Station}, which is very close to the port between Shenzhen and Hong Kong. We can see these areas have a very similar characteristic: there are many residents, {\it e.g.}, workers, students, employees, or passengers, visiting to these areas every day.

Compared with the infection number based method, the D$^2$PRI network based method detected more key areas, and the importance distinctions between urban zones were more significant. Based on the knowledge offered by Fig.~\ref{fig:pagerank}, the Shenzhen government allocated more health resources to the key areas to prevent and control epidemic outbreaks.

\section{Related works}

{\bf Epidemic Propagation:}  In the literature, epidemic propagation models could be divided as three classes: compartment models, network epidemiology models, and metapopulation models~\cite{metapopulation}. The compartment model~\cite{sir} is the simplest epidemic model, which assume all individuals in a single population and have the same probability to contact each others. It is suitable for epidemic propagation in ``well-mixed'' populations, such as smallpox in a village of a developing country~\cite{smallpox}. The network epidemiology models assume individuals in a single population are connected by an underlying network. The disease propagates through network edges. Empirical works of the network epidemiology models such as transmission of HIV/ADIS over a sexual relationships network~\cite{hiv}. Limited by the issue of network complexity and availability, very few works use network epidemiology models to analyze large spatial scale epidemic propagation.


The metapopulation network model adopted by this paper is designed for analyzing dynamics of spatially separated populations with interactions. One kind of work in this model is using empirical network data to analyze disease outbreaks in real world. For example, using global aviation networks to study outbreaks of SARS and H1N1~\cite{sars,brockmann2013hidden}, and using mobile phone data to analyze Malaria propagation in Kenya~\cite{malaria}. The other kind is to study the dynamic laws of the metapopulation network, such as the Zipf's law and the Heaps' law~\cite{wang2011evolution}. To the best of our knowledge, this paper is the first work that studying the network inference problem for metapopulation models. Besides, most of empirical works of metapopulation focus on inter-city disease propagation. This paper is also the first empirical intra-city epidemic propagation work using the metapopulation model.


{\bf Network Inference:} The Network Inference problem refers to recovering the edges of an unknown network from the observations of cascades propagating over the network. The most widely used network inference framework is first proposed by~\cite{NETINF}, in which state propagation is modeled as generative probabilistic model. Many improved methods are proposed to extend the framework, such as NETRATE~\cite{NETRATE},  ConNIe~\cite{ConNIe} and etc~\cite{du2012learning}. However, most of the existed network inference methods are designed for single population scenario, where network nodes are used to denote individuals, and the state of network nodes are discrete or binary, e.g. infected or uninfected. Therefore, we can not use these network inference methods in the metapopulation network.

{\bf Urban Computing: } This paper also have relations with urban computing~\cite{urbancomputing}. In this area, research works related to our study include: data-driven urban analysis~\cite{wang5,wang3}, resident behavior prediction~\cite{wang2,wang4,wang6}, and urban safety~\cite{wang1}. To our best knowledge, our work is the earliest studies in urban computing area that try to study the urban epidemic propagation issue.


\section{Conclusions}

In this paper, a metapopulation based epidemic propagation model not requiring detailed resident mobility data was proposed. The performance of the proposed model has been verified over an empirical data set collected from a metropolis with a population of 11 million. The performances showed that the proposed method can accurately infer the underlying sub-population network and predict a disease outbreak with  567 thousand infected persons. Our model has also been adopted by the metropolis for key areas detection.





\bibliographystyle{ACM-Reference-Format}
\bibliography{sigproc}

\end{document}